\documentclass[%
 reprint,
 amsmath,amssymb,
 prl,
]{revtex4-2}

\usepackage{graphicx}
\usepackage{dcolumn}
\usepackage{bm}
\usepackage{cases}

\usepackage{blindtext}

\begin{document}


\title{Quasiparticle nature of the Bose polaron at finite temperature}
\author{Gerard Pascual}
\author{Jordi Boronat}%
\affiliation{%
 Departament de Física, Campus Nord B4-B5, Universitat Politècnica de Catalunya, 
E-08034 Barcelona, Spain
}%

\date{\today}

\begin{abstract}
The Bose polaron has attracted theoretical and experimental interest because the mobile impurity is surrounded by a bath that undergoes a superfluid-to-normal phase transition. Although many theoretical works have studied this system in its ground state, only few analyze its behavior at finite temperature. We have studied the effect of temperature on a Bose polaron system performing ab-initio Path Integral Monte Carlo simulations. This method is able to approach the critical temperature without losing accuracy, in contrast with perturbative approximations. We have calculated the polaron energy for the repulsive and attractive branches and we have observed an asymmetric behavior between the two branches. When the potential is repulsive, the polaron energy decreases when the temperature increases, and contrariwise for the attractive branch. Our results for the effective mass and the dynamical structure factor of the polaron show unambiguously that its quasiparticle nature disappears close to the critical temperature, in agreement with recent experimental findings. Finally, we have also estimated the fraction of bosons in the condensate as well as the superfluid fraction, and we have concluded that the impurity hinders the condensation of the rest of bosons.

\end{abstract}

\maketitle


Since the first experimental observations of the Bose Einstein condensate (BEC) in alkali gases \cite{Anderson1995,Ketterle95}, BEC systems have been widely used as quantum simulators of a rich variety of problems in solid and condensed matter physics. Probably, the realization of the BCS-BEC crossover in Fermi gases \cite{Giorgini2008} and the implementation of the celebrated Bose-Hubbard model are among the most relevant achievements \cite{Lewenstein2012}. All that has been possible due to the high tunability of the interatomic interactions and the versatility to explore physics in reduced geometry.  

The polaron problem, i.e.,  a single impurity immersed in a Fermi or Bose bath, is another example of the power of ultracold quantum gases to tackle old theoretical issues. The first theoretical formulation of the problem backs to Landau and Pekar~\cite{Landau1948} who studied the quantum nature of a single electron in a polar solid. The polaron couples to the medium forming a dressed particle who was termed as a quasiparticle. This quasiparticle behaves at low momenta as a free particle, but with an effective mass which includes at first order the interactions with the surrounding bath. 

Depending on the quantum statistics of the bath one distinguishes between the Fermi and Bose polaron. The Fermi polaron was studied the first due to the experimental achievement of Feshbach resonances that made possible to observe its properties \cite{Scazza2017,Adlong2020}. More recently, the Bose polaron has attracted also the interest from both theory and experiment. In fact, the Bose polaron offers the stimulating opportunity of studying the physics of the impurity in a bath that suffers a phase transition from a superfluid to a normal gas at a critical temperature $T_c$. Interestingly, it has been theoretically predicted that the Bose polaron loses its quasiparticle nature at the critical temperature of the bath.    

In a first approach, the Hamiltonian of a mobile impurity in a bath of bosons can be simplified using the Bogoliubov approximation. The resulting Hamiltonian can then be mapped onto the Frölich polaron Hamiltonian and solved using variational schemes~\cite{Cucchi2006,Temp2009}. However, this theory is only valid for weak boson-impurity interactions where the Bogoliubov approximation still holds. Other theoretical approaches that study strong coupling systems have been considered: renormalization group theory \cite{Grusdt2017}, quantum Monte Carlo \cite{Ardila2015,Ardila2016,Ardila2020}, variational methods \cite{Li2014,Levinsen2015,Yoshida2018,Shchadilova2016,Drescher2019,Loon2018}, and diagrammatic approaches \cite{Pattrick2013}, as well as analysis on the nonequilibrium dynamics of the quasiparticle \cite{Lampo2018,Kahn2021,Mistakidis2019,Mistakidis2020,Mistakidis2021}. Nonetheless, the effect of temperature has not been included in such analysis. Experimental studies have also been performed in the regime of strong boson-impurity interaction \cite{Jorgensen2016,Hu2016} concluding that a study on the effects of temperature on this system would be interesting.

Only few works included temperature to the system using: perturbation theory \cite{Levinsen2017} (only valid for a weak boson-impurity interaction), time-dependent Hartree-Fock Bogoliubov (TDHFB) theory \cite{Boudjemaa2014_I,Boudjemaa2014_II} and large $1/N$ expansions \cite{Pastukhov2018}. Other works that proposed a diagrammatic scheme and studied the strong coupling regime have found unusual behaviors such as the splitting of the  ground state quasiparticle into two branches when the temperature is increased \cite{Guenther2018}. Nevertheless, a more recent work claimed that indeed the number of attractive branches is related to the number of hole excitations in the proposed ansatzes \cite{Field2020}. A recent study using dynamical variational theory found broadening of the spectral lines with increasing temperature~\cite{Dzsotjan2020}.  Besides, recent experiments that have performed temperature analysis on the Bose polaron have concluded that, close to $T_c$, the quasiparticle picture disappears \cite{Yan2020}. This breakdown of the quasiparticle has been theoretically suggested in Ref.~\cite{Sachdev2011} and  seen in cuprate superconductors \cite{Lee2006,Hartnoll2015}.

In this Letter, we characterize in an exact way (within controlled statistical noise) the Bose polaron at finite temperatures performing ab initio Path Integral Monte Carlo (PIMC) simulations. Contrarily to previous perturbative and variational approaches, the PIMC conserves accuracy when $T_c$  is approached and crossed. Our results agree with experimental signatures~\cite{Yan2020} on the suppression of the quasiparticle behavior of the polaron when the critical temperature is approached. The dynamic structure factor of the polaron does show a single peak with an energy increasing as the squared momenta, in contrast with previous perturbative results which show a double structure until $T_c$.

\textit{Model}.---The Hamiltonian of a mobile impurity surrounded by a bath of $N$ bosons at temperature $T$ is described as follows, 
\begin{equation}\label{eq:Hamiltonian}
    H = -\sum_{i=1}^{N}\nabla^2_i - \nabla^2_I
        + \sum_{i<j} V_B(r_{ij}) +\sum_{i=1}^N V_I(r_{iI})\ ,
\end{equation}
written in units of $\hbar^2/(2m_B)$ where we consider the mass of the bosons ($m_B$) and of the impurity ($m_I$) equal. The two first terms of the Hamiltonian correspond to the kinetic energy of the bosons and the impurity. The rest of the terms are the boson-boson ($V_B$) and the boson-impurity ($V_I$) interactions, that depend only on the modulus of the interparticle distance. We model the different potentials using continuous functions~\cite{Supplement}: $V_B(r)$ is always a repulsive potential whereas $V_I(r)$ is chosen as repulsive or attractive depending on the repulsive or attractive polaron branch under study, respectively. Our study is carried out under universality conditions for the gas parameter $na^3$ \cite{Giorgini1999}, with $n=N/V$ the density of the bath and $a$ the s-wave scattering length of the B-B interactions (considered as unit length). In all our work, we use as energy unit $g n = (4 \pi \hbar^2/ m_Ba^2)\, na^3$, with $g$ the mean-field strength interaction.

\begin{figure}
	\centering
	\includegraphics[width=8.6cm,height=4cm]{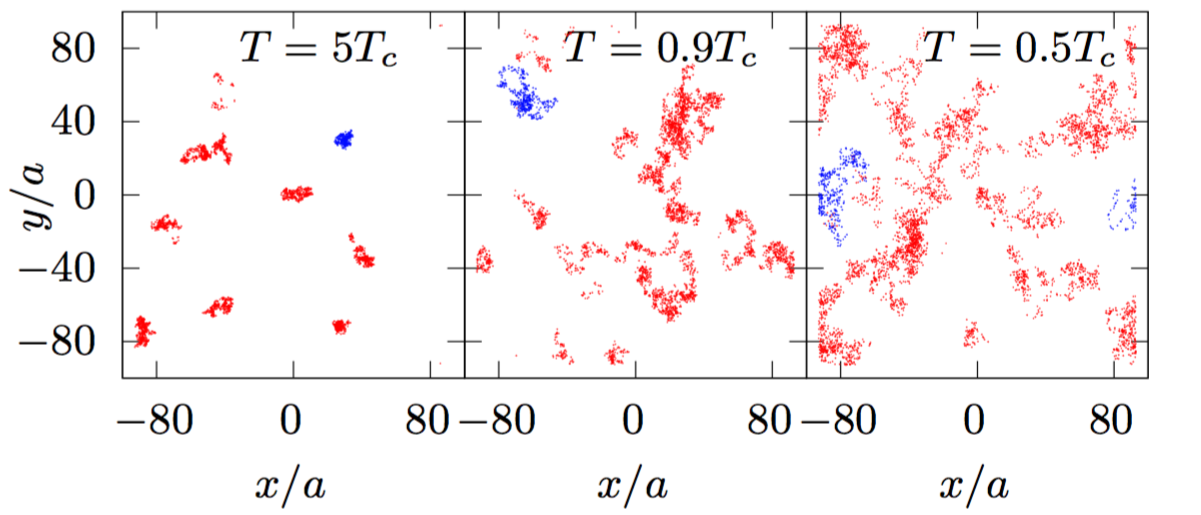}
	\caption{\label{fig:positions} Projections of the positions of the particles in 2D at different temperatures. Dots in blue represent the impurity.}
\end{figure}

\textit{Method}.---The properties of the system, described by the Hamiltonian in Eq. \ref{eq:Hamiltonian}, are calculated by means of PIMC simulations. Not only does this method allow us to calculate fundamental properties of the system such as the energy, but it gives us also access to structural information, namely the effective mass, the radial distribution function, the one-body density matrix, and the dynamic structure factor. Furthermore, no need of approximations is required in this technique if the number imaginary-time steps (\textit{beads}) in each particle (\textit{polymer}) is sufficiently large. Indeed, as the temperature decreases, a larger number of \textit{beads} is required since \textit{polymers}, i.e., particles, become more delocalized (see Fig. \ref{fig:positions}). In order to reduce the number of beads to a manageable level, we use the fourth-order Chin action that can be even sixth-order in an effective way \cite{Sakkos2009}. Finally, we use the worm algorithm to sample the permutation space (see further technical details in Ref.~\cite{Supplement}). 

\textit{Results}.--- We have performed PIMC simulations for a low density Bose 
gas ($na^3 = 10^{-4}$). At such low densities, the universality in terms of the 
gas parameter holds \cite{Giorgini1999} and the system only depends on the ratio of the scattering 
lengths $a/b$, with $b$ the s-wave scattering length of the polaron-bath 
interaction. We start computing general properties when the system is thermalized.

The optimal number of beads at each temperature is determined by studying the convergence of the energy when this number is progressively incremented\cite{Sakkos2009}. The computational cost increases dramatically when the temperature approaches the zero limit since the number of beads is inversely proportional to the temperature. In fact, only the use of high-order actions like the one used here allows for reliable calculations at temperatures as low as $\sim 0.2 \, T_c$. 

\begin{figure}[b]
	\centering
	\includegraphics[width=8.6cm,height=10cm]{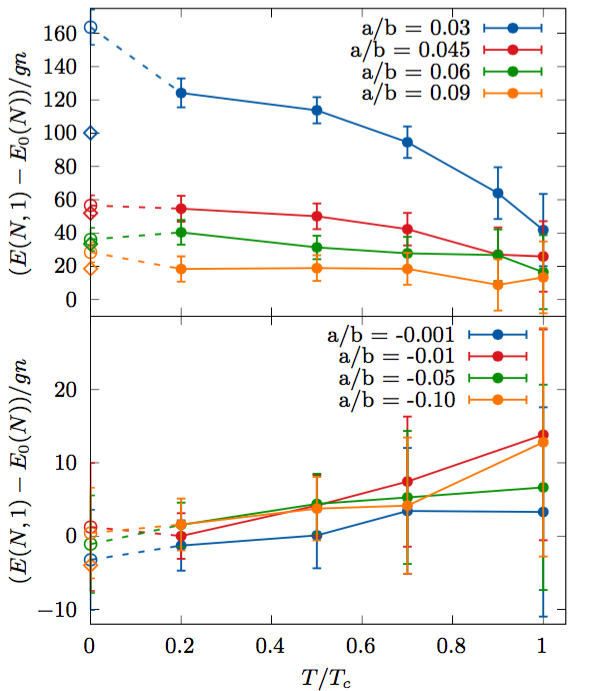}
	\caption{\label{fig:T} \textit{Top box}: Polaron energy as a function of temperature for the repulsive branch ($na^3 = 10^{-4}$). \textit{Bottom box}: Same for the attractive branch ($na^3 = 10^{-4}$). Empty circles correspond to data computed using PIGS method and the diamonds correspond to the mean field approach (Eq. \ref{eq:MeanField}).}
\end{figure}

The simulations for the bath are carried out with $N=64$ bosons using periodic boundary conditions. The volume of the simulation box is such that $na^3=10^{-4}$. A second set of simulations is performed including a mobile impurity (polaron system), keeping the volume constant. The difference between the total energy of the bath with polaron and bath systems  is termed the polaron energy, $E(N,1)-E_{0}(N)$, which is in fact the chemical potential of the polaron. Fig. \ref{fig:T} shows the PIMC results of the polaron energy at different temperature and scattering length ratio in both the attractive ($b<0$) and repulsive ($b>0$) branches. The PIMC method cannot be extended to the $T=0$ limit but to know the ground-state values is also interesting. To this end, we used the zero-temperature version of PIMC, known as th Path Integral Ground State (PIGS) method \cite{Supplement}. The polaron energies at $T=0$ are shown as empty circles in Fig.~\ref{fig:T}. We compared our results at $T=0$ with the predictions of mean-field theory up to second order (in units of $gn$)   \cite{Temp2009}, 

\begin{equation}\label{eq:MeanField}
    E(N,1)-E_{0}(N) = \frac{b}{a} + \frac{32}{3\sqrt{\pi}}\sqrt{na^3}\frac{b^2}{a^2} \ .
\end{equation}

The mean-field results (\ref{eq:MeanField}), which hold only for weak interactions \cite{Ardila2015}, are shown as  empty diamonds in Fig. \ref{fig:T}. Notice that for the attractive branch we have only compared the mean-field result for $a/b = -0.1$ as, below that value,  the mean-field approach is no longer valid \cite{Ardila2015}. 

We tested different model potentials for the attractive and the repulsive branches and no differences on the results were observed. We also explored the polaron system at different densities in order to compare our results with other works. However, if we increase the density in excess, the shape of the interatomic potentials starts becoming relevant and the universality in terms of the gas parameter is lost. On the other hand, if we decrease more the density, the statistical noise of the sampling starts growing and it becomes of the same order of the polaron energy. A similar effect appears when we simulate the system at temperatures above $T_c$. At such temperatures, where there is not a condensate of bosons in the bath, the thermal noise suppresses practically all the signal of the energy difference between the impurity and the bath systems. Furthermore, we also simulated the system increasing the 
number of particles in the bath  ($N=128$)  without seeing significant changes in the results. 

For the repulsive branch, we observe that the polaron energy decreases when the temperature increases. This indicates that, when the fraction of particles in the condensate is large, adding an impurity into the system requires more energy. Noticeably, in the attractive branch we see the opposite behavior: the polaron energy increases with temperature. For weak attraction, our results agree qualitatively with experiments \cite{Yan2020} but we do not obtain the experimental change of tendency observed for stronger interactions ($a/b\approx -0.01$) \cite{Yan2020}.

One of the most clear signatures of the quasiparticle nature of the polaron is its effective mass, which characterizes its excitation spectrum at low momenta, $\epsilon(q)= \hbar^2 q^2/2 m^*$. In QMC methods, the effective mass of the 
polaron can be obtained from its diffusion coefficient in imaginary-time, 
\begin{equation}
    \frac{m_I}{m^*} = \lim_{\tau \rightarrow \infty} \frac{\big<|\Delta 
\vec{r_I} (\tau)|^2 \big>}{6D\tau} \ ,
\label{effectivemass}
\end{equation}
where $m^*$ is the effective mass, $\tau$ is the imaginary time, $D = \hbar^2/(2m_I)$, and $\Delta \vec{r_I}(\tau) = \vec{r_I}(\tau)-\vec{r_I}(0)$ is the displacement of the impurity in imaginary time \cite{Boninsegni1995, Boronat1999}. The brackets indicate the mean over the samples.

\begin{figure}
	\centering
	\includegraphics[width=8.6cm,height=3.2cm]{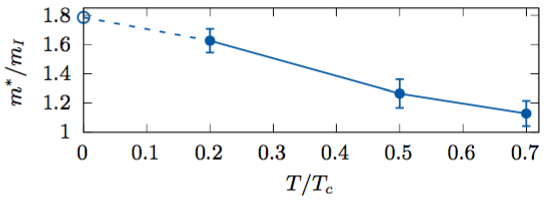}
	\caption{\label{fig:effmass} Effective mass of the Bose polaron as a 
function of temperature for a system with $a/b = 0.06$ and $na^3 = 10^{-5}$. The 
empty circle corresponds to data computed using quantum Monte Carlo methods at 
$T=0T_c$ \cite{Ardila2015}.}
\end{figure}

With the PIMC method we can only compute displacements up to $\tau = \beta/2$ where $\beta = 1/(k_B T)$ and $k_B$ is the Boltzmann constant. To overcome this limit and approach large enough imaginary times, we sample the impurity \textit{polymer} when it is open. Fig. \ref{fig:effmass} shows the effect of the temperature on the effective mass, calculated for the repulsive branch. As one can see, close the critical temperature the effective mass ($m^*/m_I$) tends to $1$. The effective mass decreases with $T$ because the increase of the kinetic energy of particles in the bath reduces the size of the \textit{hole} created by the polaron. This result reinforces the idea that close to the critical temperature the quasiparticle picture of the polaron vanishes \cite{Yan2020}. 

\begin{figure}[h!]
	\centering
	\includegraphics[width=8.6cm,height=3.6cm]{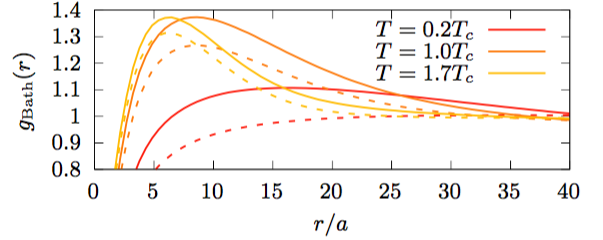}
	\caption{\label{fig:gr} Radial distribution function of bath-bath particles at different temperatures ($a/b = 0.06$ and $na^3 = 10^{-4}$). Dashed lines correspond to systems without the impurity.}
\end{figure}

It is interesting to study the influence of the polaron in the properties of the Bose bath. To this end, we calculated the two-body distribution function of the bath particles, $g_{\text{Bath}}(r)$.
Fig. \ref{fig:gr} shows these results as a function of temperature and it compares them with the system without the impurity (dashed lines). We see that, in general, temperature adds kinetic energy to the system and, thus, particles in the bath can overcome the repulsive potential and approach each other. When an impurity is added to the system we notice an effect of compression in the bath, probably due to the repulsive interaction between the polaron and the medium. In all cases, one can see the bunching effect in $g_{\text{Bath}}$ when the temperature increases, a reminiscent of the well-known ideal Bose gas behavior.

\begin{figure}
	\centering
	\includegraphics[width=8.6cm,height=5.2cm]{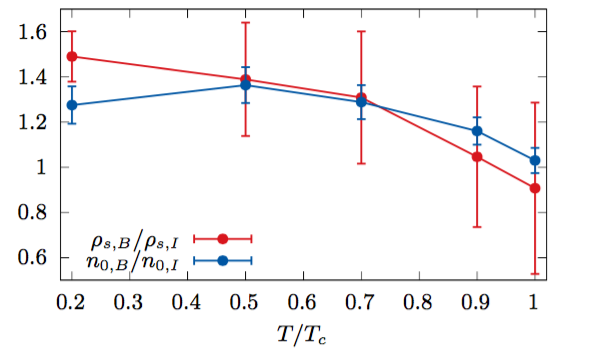}
	\caption{\label{fig:drave} \textit{Blue}: Ratio of the fraction of condensed bosons between a bath system ($n_{0,B}$) and an impurity system ($n_{0,I}$) at $a/b = 0.03$ and $na^3 = 10^{-4}$. \textit{Red}: Ratio of the superfluid density between a bath system ($\rho_{s,B}$) and an impurity system ($\rho_{s,I}$) at the same conditions.}
\end{figure}

Off-diagonal long-range order and superfluidity of the bath are also affected by the presence of the impurity.
To quantify this effect, we calculated first the ratio of the fraction of particles in the Bose-Einstein condensate between the bath and the impurity systems ($n_{0,B}/n_{0,I}$). Fig. \ref{fig:drave} shows the ratio of the systems without and with the impurity and its dependency on temperature. We see that in the bath system there are more bosons in the condensate than in the system with the impurity. This indicates that the impurity-bath interaction promotes particles out of the condensate. Moreover, we observe that as the temperature decreases this effect becomes more evident. One expects that a similar feature can happen for the superfluid density of the bath. To corroborate this idea, we calculated the winding number in order to infer the superfluid density \cite{Ceperley1995}. The results are also plotted in Fig. \ref{fig:drave} and they show that, in the same way as with the fraction of particles in the condensate, the superfluid density decreases when the polaron is present and that the fraction of the superfluid densities, $\rho_{s,B}/\rho_{s,I}$, diminishes with temperature.

\begin{figure*}
    \centering
    \includegraphics[width=18cm,height=6cm]{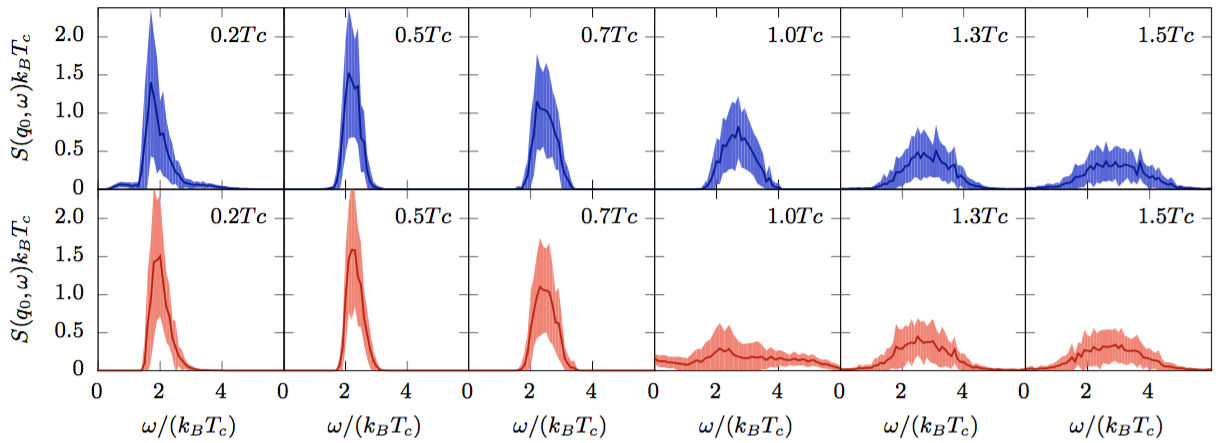}
    \caption{\label{fig:sktau} Dynamic structure factor as a function of temperature at $q_0a = 0.16$. The blue lines correspond to repulsive interaction ($a/b = 0.03$) and the red lines correspond to the attractive interaction ($a/b = -0.01$). The dark lines show the mean value and the shaded regions plot the error.}
\end{figure*}

Finally we calculated the dynamic structure factor $S(q,\omega)$ from the intermediate scattering function $F(q,\tau) = \langle e^{iqr_I(\tau)}e^{-iqr_I(0)}\rangle$ (where the brackets stand for the thermal average) through the inverse Laplace transform. This inverse Laplace transform is computed optimizing the parameters of a model function $\tilde{S}(q,\omega)$ using the simulated annealing technique \cite{Ferre2016,MacKeow1997}. Figure \ref{fig:sktau} shows the dynamic structure factor for the repulsive and attractive branches as a function of temperature. As one can see, the errorbars are quite large due to two features. On one side, due to the poor statistics got in the sampling of  $F(q,\tau)$ as it is an operator of order one. On the other, due to the uncertainties in the calculation of the ill-posed inverse Laplace transform. Anyway, our results show clearly how the quasiparticle behavior of the excited polaron is lost when $T_c$ is crossed, in agreement with experimental observations \cite{Yan2020}. Also, our results do not show any signature of a second peak in 
$S(q,\omega)$ as initial perturbative estimations claimed \cite{Guenther2018}. Similar results to the ones obtained here appeared time ago for the problem of a single $^3$He impurity in a $^4$He bath \cite{Fabrocini1986, Fabrocini1998}.

\textit{Conclusions}.--- The main purpose of this Letter was to study the effect of temperature (up to $T_c$) on a Bose polaron system in a similar way as recent experiments have done \cite{Yan2020}. On the basis of the results obtained, we see similar trends for weak interactions in the attractive branch and also a particular asymmetry between the attractive and the repulsive branches. We do not observe a significant change in the attractive branch at various scattering length ratios as Ref. \cite{Yan2020} suggests. However, we agree that the quasiparticle picture vanishes close to the critical temperature since we see that the effective mass tends to the bare mass  when the temperature increases. Furthermore, we also notice that the impurity disturbs the bath and reduces the number of bosons in the condensate as well as the superfluid density. This effect is in agreement with the behavior of the polaron energy for the repulsive branch where we observe that, at low temperatures, adding an impurity requires more energy. Indeed, this increase in the polaron energy as temperature decreases can be understood as the competition between the bath, that tries to condensate all the particles, and the impurity that (slightly) hinders this condensation due to its interaction with the bath. As it is clear from our results for the dynamic structure factor, the polaron loses dramatically its quasiparticle nature when the critical temperature is crossed. 

Currently, only one experiment studied the Bose polaron with an attractive potential \cite{Yan2020}. This work suggests that it would be interesting to conduct more experimental research analyzing the effect of temperature for repulsive potentials as well. Finally, our work could be easily extended to low dimensions by constraining the sampling of the PIMC. 

\textit{Acknowledgments}.--- This work has been supported by the Ministerio de Economia, Industria y Competitividad (MINECO, Spain) under grant
        No. FIS2017-84114-C2-1-P.   We also acknowledge financial support from Secretaria d'Universitats i Recerca del Departament d'Empresa i Coneixement de la Generalitat de Catalunya, co-funded by the European Union Regional Development Fund within the ERDF Operational Program of Catalunya (project QuantumCat, ref. 001-P-001644).x


\bibliography{main}

\begin{thebibliography}{58}%
\makeatletter
\providecommand \@ifxundefined [1]{%
 \@ifx{#1\undefined}
}%
\providecommand \@ifnum [1]{%
 \ifnum #1\expandafter \@firstoftwo
 \else \expandafter \@secondoftwo
 \fi
}%
\providecommand \@ifx [1]{%
 \ifx #1\expandafter \@firstoftwo
 \else \expandafter \@secondoftwo
 \fi
}%
\providecommand \natexlab [1]{#1}%
\providecommand \enquote  [1]{``#1''}%
\providecommand \bibnamefont  [1]{#1}%
\providecommand \bibfnamefont [1]{#1}%
\providecommand \citenamefont [1]{#1}%
\providecommand \href@noop [0]{\@secondoftwo}%
\providecommand \href [0]{\begingroup \@sanitize@url \@href}%
\providecommand \@href[1]{\@@startlink{#1}\@@href}%
\providecommand \@@href[1]{\endgroup#1\@@endlink}%
\providecommand \@sanitize@url [0]{\catcode `\\12\catcode `\$12\catcode
  `\&12\catcode `\#12\catcode `\^12\catcode `\_12\catcode `\%12\relax}%
\providecommand \@@startlink[1]{}%
\providecommand \@@endlink[0]{}%
\providecommand \url  [0]{\begingroup\@sanitize@url \@url }%
\providecommand \@url [1]{\endgroup\@href {#1}{\urlprefix }}%
\providecommand \urlprefix  [0]{URL }%
\providecommand \Eprint [0]{\href }%
\providecommand \doibase [0]{https://doi.org/}%
\providecommand \selectlanguage [0]{\@gobble}%
\providecommand \bibinfo  [0]{\@secondoftwo}%
\providecommand \bibfield  [0]{\@secondoftwo}%
\providecommand \translation [1]{[#1]}%
\providecommand \BibitemOpen [0]{}%
\providecommand \bibitemStop [0]{}%
\providecommand \bibitemNoStop [0]{.\EOS\space}%
\providecommand \EOS [0]{\spacefactor3000\relax}%
\providecommand \BibitemShut  [1]{\csname bibitem#1\endcsname}%
\let\auto@bib@innerbib\@empty
\bibitem [{\citenamefont {Anderson}\ \emph {et~al.}(1995)\citenamefont
  {Anderson}, \citenamefont {Ensher}, \citenamefont {Matthews}, \citenamefont
  {Wieman},\ and\ \citenamefont {Cornell}}]{Anderson1995}%
  \BibitemOpen
  \bibfield  {author} {\bibinfo {author} {\bibfnamefont {M.~H.}\ \bibnamefont
  {Anderson}}, \bibinfo {author} {\bibfnamefont {J.~R.}\ \bibnamefont
  {Ensher}}, \bibinfo {author} {\bibfnamefont {M.~R.}\ \bibnamefont
  {Matthews}}, \bibinfo {author} {\bibfnamefont {C.~E.}\ \bibnamefont
  {Wieman}},\ and\ \bibinfo {author} {\bibfnamefont {E.~A.}\ \bibnamefont
  {Cornell}},\ }\bibfield  {title} {\bibinfo {title} {Observation of
  bose-einstein condensation in a dilute atomic vapor},\ }\href
  {https://doi.org/10.1126/science.269.5221.198} {\bibfield  {journal}
  {\bibinfo  {journal} {Science}\ }\textbf {\bibinfo {volume} {269}},\ \bibinfo
  {pages} {198} (\bibinfo {year} {1995})}\BibitemShut {NoStop}%
\bibitem [{\citenamefont {Davis}\ \emph {et~al.}(1995)\citenamefont {Davis},
  \citenamefont {Mewes}, \citenamefont {Andrews}, \citenamefont {van Druten},
  \citenamefont {Durfee}, \citenamefont {Kurn},\ and\ \citenamefont
  {Ketterle}}]{Ketterle95}%
  \BibitemOpen
  \bibfield  {author} {\bibinfo {author} {\bibfnamefont {K.~B.}\ \bibnamefont
  {Davis}}, \bibinfo {author} {\bibfnamefont {M.~O.}\ \bibnamefont {Mewes}},
  \bibinfo {author} {\bibfnamefont {M.~R.}\ \bibnamefont {Andrews}}, \bibinfo
  {author} {\bibfnamefont {N.~J.}\ \bibnamefont {van Druten}}, \bibinfo
  {author} {\bibfnamefont {D.~S.}\ \bibnamefont {Durfee}}, \bibinfo {author}
  {\bibfnamefont {D.~M.}\ \bibnamefont {Kurn}},\ and\ \bibinfo {author}
  {\bibfnamefont {W.}~\bibnamefont {Ketterle}},\ }\bibfield  {title} {\bibinfo
  {title} {Bose-einstein condensation in a gas of sodium atoms},\ }\href
  {https://doi.org/10.1103/PhysRevLett.75.3969} {\bibfield  {journal} {\bibinfo
   {journal} {Phys. Rev. Lett.}\ }\textbf {\bibinfo {volume} {75}},\ \bibinfo
  {pages} {3969} (\bibinfo {year} {1995})}\BibitemShut {NoStop}%
\bibitem [{\citenamefont {Giorgini}\ \emph {et~al.}(2008)\citenamefont
  {Giorgini}, \citenamefont {Pitaevskii},\ and\ \citenamefont
  {Stringari}}]{Giorgini2008}%
  \BibitemOpen
  \bibfield  {author} {\bibinfo {author} {\bibfnamefont {S.}~\bibnamefont
  {Giorgini}}, \bibinfo {author} {\bibfnamefont {L.~P.}\ \bibnamefont
  {Pitaevskii}},\ and\ \bibinfo {author} {\bibfnamefont {S.}~\bibnamefont
  {Stringari}},\ }\bibfield  {title} {\bibinfo {title} {Theory of ultracold
  atomic fermi gases},\ }\href {https://doi.org/10.1103/RevModPhys.80.1215}
  {\bibfield  {journal} {\bibinfo  {journal} {Rev. Mod. Phys.}\ }\textbf
  {\bibinfo {volume} {80}},\ \bibinfo {pages} {1215} (\bibinfo {year}
  {2008})}\BibitemShut {NoStop}%
\bibitem [{\citenamefont {Lewenstein}\ \emph {et~al.}(2012)\citenamefont
  {Lewenstein}, \citenamefont {Sanpera},\ and\ \citenamefont
  {Ahufinger}}]{Lewenstein2012}%
  \BibitemOpen
  \bibfield  {author} {\bibinfo {author} {\bibfnamefont {M.}~\bibnamefont
  {Lewenstein}}, \bibinfo {author} {\bibfnamefont {A.}~\bibnamefont
  {Sanpera}},\ and\ \bibinfo {author} {\bibfnamefont {V.}~\bibnamefont
  {Ahufinger}},\ }\href {https://books.google.es/books?id=Wpl91RDxV5IC} {\emph
  {\bibinfo {title} {Ultracold Atoms in Optical Lattices: Simulating quantum
  many-body systems}}}\ (\bibinfo  {publisher} {OUP Oxford},\ \bibinfo {year}
  {2012})\BibitemShut {NoStop}%
\bibitem [{\citenamefont {Landau}\ and\ \citenamefont
  {Pekar}(1948)}]{Landau1948}%
  \BibitemOpen
  \bibfield  {author} {\bibinfo {author} {\bibfnamefont {L.~D.}\ \bibnamefont
  {Landau}}\ and\ \bibinfo {author} {\bibfnamefont {S.~I.}\ \bibnamefont
  {Pekar}},\ }\bibfield  {title} {\bibinfo {title} {Effective mass of a
  polaron},\ }\href@noop {} {\bibfield  {journal} {\bibinfo  {journal} {Zh.
  Eksp. Teor. Fiz}\ }\textbf {\bibinfo {volume} {18}},\ \bibinfo {pages}
  {419–423} (\bibinfo {year} {1948})}\BibitemShut {NoStop}%
\bibitem [{\citenamefont {Scazza}\ \emph {et~al.}(2017)\citenamefont {Scazza},
  \citenamefont {Valtolina}, \citenamefont {Massignan}, \citenamefont {Recati},
  \citenamefont {Amico}, \citenamefont {Burchianti}, \citenamefont {Fort},
  \citenamefont {Inguscio}, \citenamefont {Zaccanti},\ and\ \citenamefont
  {Roati}}]{Scazza2017}%
  \BibitemOpen
  \bibfield  {author} {\bibinfo {author} {\bibfnamefont {F.}~\bibnamefont
  {Scazza}}, \bibinfo {author} {\bibfnamefont {G.}~\bibnamefont {Valtolina}},
  \bibinfo {author} {\bibfnamefont {P.}~\bibnamefont {Massignan}}, \bibinfo
  {author} {\bibfnamefont {A.}~\bibnamefont {Recati}}, \bibinfo {author}
  {\bibfnamefont {A.}~\bibnamefont {Amico}}, \bibinfo {author} {\bibfnamefont
  {A.}~\bibnamefont {Burchianti}}, \bibinfo {author} {\bibfnamefont
  {C.}~\bibnamefont {Fort}}, \bibinfo {author} {\bibfnamefont {M.}~\bibnamefont
  {Inguscio}}, \bibinfo {author} {\bibfnamefont {M.}~\bibnamefont {Zaccanti}},\
  and\ \bibinfo {author} {\bibfnamefont {G.}~\bibnamefont {Roati}},\ }\bibfield
   {title} {\bibinfo {title} {Repulsive fermi polarons in a resonant mixture of
  ultracold $^{6}\mathrm{Li}$ atoms},\ }\href
  {https://doi.org/10.1103/PhysRevLett.118.083602} {\bibfield  {journal}
  {\bibinfo  {journal} {Phys. Rev. Lett.}\ }\textbf {\bibinfo {volume} {118}},\
  \bibinfo {pages} {083602} (\bibinfo {year} {2017})}\BibitemShut {NoStop}%
\bibitem [{\citenamefont {Adlong}\ \emph {et~al.}(2020)\citenamefont {Adlong},
  \citenamefont {Liu}, \citenamefont {Scazza}, \citenamefont {Zaccanti},
  \citenamefont {Oppong}, \citenamefont {F\"olling}, \citenamefont {Parish},\
  and\ \citenamefont {Levinsen}}]{Adlong2020}%
  \BibitemOpen
  \bibfield  {author} {\bibinfo {author} {\bibfnamefont {H.~S.}\ \bibnamefont
  {Adlong}}, \bibinfo {author} {\bibfnamefont {W.~E.}\ \bibnamefont {Liu}},
  \bibinfo {author} {\bibfnamefont {F.}~\bibnamefont {Scazza}}, \bibinfo
  {author} {\bibfnamefont {M.}~\bibnamefont {Zaccanti}}, \bibinfo {author}
  {\bibfnamefont {N.~D.}\ \bibnamefont {Oppong}}, \bibinfo {author}
  {\bibfnamefont {S.}~\bibnamefont {F\"olling}}, \bibinfo {author}
  {\bibfnamefont {M.~M.}\ \bibnamefont {Parish}},\ and\ \bibinfo {author}
  {\bibfnamefont {J.}~\bibnamefont {Levinsen}},\ }\bibfield  {title} {\bibinfo
  {title} {Quasiparticle lifetime of the repulsive fermi polaron},\ }\href
  {https://doi.org/10.1103/PhysRevLett.125.133401} {\bibfield  {journal}
  {\bibinfo  {journal} {Phys. Rev. Lett.}\ }\textbf {\bibinfo {volume} {125}},\
  \bibinfo {pages} {133401} (\bibinfo {year} {2020})}\BibitemShut {NoStop}%
\bibitem [{\citenamefont {Cucchietti}\ and\ \citenamefont
  {Timmermans}(2006)}]{Cucchi2006}%
  \BibitemOpen
  \bibfield  {author} {\bibinfo {author} {\bibfnamefont {F.~M.}\ \bibnamefont
  {Cucchietti}}\ and\ \bibinfo {author} {\bibfnamefont {E.}~\bibnamefont
  {Timmermans}},\ }\bibfield  {title} {\bibinfo {title} {Strong-coupling
  polarons in dilute gas bose-einstein condensates},\ }\href
  {https://doi.org/10.1103/PhysRevLett.96.210401} {\bibfield  {journal}
  {\bibinfo  {journal} {Phys. Rev. Lett.}\ }\textbf {\bibinfo {volume} {96}},\
  \bibinfo {pages} {210401} (\bibinfo {year} {2006})}\BibitemShut {NoStop}%
\bibitem [{\citenamefont {Tempere}\ \emph {et~al.}(2009)\citenamefont
  {Tempere}, \citenamefont {Casteels}, \citenamefont {Oberthaler},
  \citenamefont {Knoop}, \citenamefont {Timmermans},\ and\ \citenamefont
  {Devreese}}]{Temp2009}%
  \BibitemOpen
  \bibfield  {author} {\bibinfo {author} {\bibfnamefont {J.}~\bibnamefont
  {Tempere}}, \bibinfo {author} {\bibfnamefont {W.}~\bibnamefont {Casteels}},
  \bibinfo {author} {\bibfnamefont {M.~K.}\ \bibnamefont {Oberthaler}},
  \bibinfo {author} {\bibfnamefont {S.}~\bibnamefont {Knoop}}, \bibinfo
  {author} {\bibfnamefont {E.}~\bibnamefont {Timmermans}},\ and\ \bibinfo
  {author} {\bibfnamefont {J.~T.}\ \bibnamefont {Devreese}},\ }\bibfield
  {title} {\bibinfo {title} {Feynman path-integral treatment of the
  bec-impurity polaron},\ }\href {https://doi.org/10.1103/PhysRevB.80.184504}
  {\bibfield  {journal} {\bibinfo  {journal} {Phys. Rev. B}\ }\textbf {\bibinfo
  {volume} {80}},\ \bibinfo {pages} {184504} (\bibinfo {year}
  {2009})}\BibitemShut {NoStop}%
\bibitem [{\citenamefont {Grusdt}\ \emph {et~al.}(2017)\citenamefont {Grusdt},
  \citenamefont {Schmidt}, \citenamefont {Shchadilova},\ and\ \citenamefont
  {Demler}}]{Grusdt2017}%
  \BibitemOpen
  \bibfield  {author} {\bibinfo {author} {\bibfnamefont {F.}~\bibnamefont
  {Grusdt}}, \bibinfo {author} {\bibfnamefont {R.}~\bibnamefont {Schmidt}},
  \bibinfo {author} {\bibfnamefont {Y.~E.}\ \bibnamefont {Shchadilova}},\ and\
  \bibinfo {author} {\bibfnamefont {E.}~\bibnamefont {Demler}},\ }\bibfield
  {title} {\bibinfo {title} {Strong-coupling bose polarons in a bose-einstein
  condensate},\ }\href {https://doi.org/10.1103/PhysRevA.96.013607} {\bibfield
  {journal} {\bibinfo  {journal} {Phys. Rev. A}\ }\textbf {\bibinfo {volume}
  {96}},\ \bibinfo {pages} {013607} (\bibinfo {year} {2017})}\BibitemShut
  {NoStop}%
\bibitem [{\citenamefont {Ardila}\ and\ \citenamefont
  {Giorgini}(2015)}]{Ardila2015}%
  \BibitemOpen
  \bibfield  {author} {\bibinfo {author} {\bibfnamefont {L.~A.~P.}\
  \bibnamefont {Ardila}}\ and\ \bibinfo {author} {\bibfnamefont
  {S.}~\bibnamefont {Giorgini}},\ }\bibfield  {title} {\bibinfo {title}
  {Impurity in a bose-einstein condensate: Study of the attractive and
  repulsive branch using quantum monte carlo methods},\ }\href
  {https://doi.org/10.1103/PhysRevA.92.033612} {\bibfield  {journal} {\bibinfo
  {journal} {Phys. Rev. A}\ }\textbf {\bibinfo {volume} {92}},\ \bibinfo
  {pages} {033612} (\bibinfo {year} {2015})}\BibitemShut {NoStop}%
\bibitem [{\citenamefont {Ardila}\ and\ \citenamefont
  {Giorgini}(2016)}]{Ardila2016}%
  \BibitemOpen
  \bibfield  {author} {\bibinfo {author} {\bibfnamefont {L.~A.~P.}\
  \bibnamefont {Ardila}}\ and\ \bibinfo {author} {\bibfnamefont
  {S.}~\bibnamefont {Giorgini}},\ }\bibfield  {title} {\bibinfo {title} {Bose
  polaron problem: Effect of mass imbalance on binding energy},\ }\href
  {https://doi.org/10.1103/PhysRevA.94.063640} {\bibfield  {journal} {\bibinfo
  {journal} {Phys. Rev. A}\ }\textbf {\bibinfo {volume} {94}},\ \bibinfo
  {pages} {063640} (\bibinfo {year} {2016})}\BibitemShut {NoStop}%
\bibitem [{\citenamefont {Ardila}\ \emph {et~al.}(2020)\citenamefont {Ardila},
  \citenamefont {Astrakharchik},\ and\ \citenamefont {Giorgini}}]{Ardila2020}%
  \BibitemOpen
  \bibfield  {author} {\bibinfo {author} {\bibfnamefont {L.~A.~P.}\
  \bibnamefont {Ardila}}, \bibinfo {author} {\bibfnamefont {G.~E.}\
  \bibnamefont {Astrakharchik}},\ and\ \bibinfo {author} {\bibfnamefont
  {S.}~\bibnamefont {Giorgini}},\ }\bibfield  {title} {\bibinfo {title} {Strong
  coupling bose polarons in a two-dimensional gas},\ }\href
  {https://doi.org/10.1103/PhysRevResearch.2.023405} {\bibfield  {journal}
  {\bibinfo  {journal} {Phys. Rev. Research}\ }\textbf {\bibinfo {volume}
  {2}},\ \bibinfo {pages} {023405} (\bibinfo {year} {2020})}\BibitemShut
  {NoStop}%
\bibitem [{\citenamefont {Li}\ and\ \citenamefont {Das~Sarma}(2014)}]{Li2014}%
  \BibitemOpen
  \bibfield  {author} {\bibinfo {author} {\bibfnamefont {W.}~\bibnamefont
  {Li}}\ and\ \bibinfo {author} {\bibfnamefont {S.}~\bibnamefont {Das~Sarma}},\
  }\bibfield  {title} {\bibinfo {title} {Variational study of polarons in
  bose-einstein condensates},\ }\href
  {https://doi.org/10.1103/PhysRevA.90.013618} {\bibfield  {journal} {\bibinfo
  {journal} {Phys. Rev. A}\ }\textbf {\bibinfo {volume} {90}},\ \bibinfo
  {pages} {013618} (\bibinfo {year} {2014})}\BibitemShut {NoStop}%
\bibitem [{\citenamefont {Levinsen}\ \emph {et~al.}(2015)\citenamefont
  {Levinsen}, \citenamefont {Parish},\ and\ \citenamefont
  {Bruun}}]{Levinsen2015}%
  \BibitemOpen
  \bibfield  {author} {\bibinfo {author} {\bibfnamefont {J.}~\bibnamefont
  {Levinsen}}, \bibinfo {author} {\bibfnamefont {M.~M.}\ \bibnamefont
  {Parish}},\ and\ \bibinfo {author} {\bibfnamefont {G.~M.}\ \bibnamefont
  {Bruun}},\ }\bibfield  {title} {\bibinfo {title} {Impurity in a bose-einstein
  condensate and the efimov effect},\ }\href
  {https://doi.org/10.1103/PhysRevLett.115.125302} {\bibfield  {journal}
  {\bibinfo  {journal} {Phys. Rev. Lett.}\ }\textbf {\bibinfo {volume} {115}},\
  \bibinfo {pages} {125302} (\bibinfo {year} {2015})}\BibitemShut {NoStop}%
\bibitem [{\citenamefont {Yoshida}\ \emph {et~al.}(2018)\citenamefont
  {Yoshida}, \citenamefont {Endo}, \citenamefont {Levinsen},\ and\
  \citenamefont {Parish}}]{Yoshida2018}%
  \BibitemOpen
  \bibfield  {author} {\bibinfo {author} {\bibfnamefont {S.~M.}\ \bibnamefont
  {Yoshida}}, \bibinfo {author} {\bibfnamefont {S.}~\bibnamefont {Endo}},
  \bibinfo {author} {\bibfnamefont {J.}~\bibnamefont {Levinsen}},\ and\
  \bibinfo {author} {\bibfnamefont {M.~M.}\ \bibnamefont {Parish}},\ }\bibfield
   {title} {\bibinfo {title} {Universality of an impurity in a bose-einstein
  condensate},\ }\href {https://doi.org/10.1103/PhysRevX.8.011024} {\bibfield
  {journal} {\bibinfo  {journal} {Phys. Rev. X}\ }\textbf {\bibinfo {volume}
  {8}},\ \bibinfo {pages} {011024} (\bibinfo {year} {2018})}\BibitemShut
  {NoStop}%
\bibitem [{\citenamefont {Shchadilova}\ \emph {et~al.}(2016)\citenamefont
  {Shchadilova}, \citenamefont {Schmidt}, \citenamefont {Grusdt},\ and\
  \citenamefont {Demler}}]{Shchadilova2016}%
  \BibitemOpen
  \bibfield  {author} {\bibinfo {author} {\bibfnamefont {Y.~E.}\ \bibnamefont
  {Shchadilova}}, \bibinfo {author} {\bibfnamefont {R.}~\bibnamefont
  {Schmidt}}, \bibinfo {author} {\bibfnamefont {F.}~\bibnamefont {Grusdt}},\
  and\ \bibinfo {author} {\bibfnamefont {E.}~\bibnamefont {Demler}},\
  }\bibfield  {title} {\bibinfo {title} {Quantum dynamics of ultracold bose
  polarons},\ }\href {https://doi.org/10.1103/PhysRevLett.117.113002}
  {\bibfield  {journal} {\bibinfo  {journal} {Phys. Rev. Lett.}\ }\textbf
  {\bibinfo {volume} {117}},\ \bibinfo {pages} {113002} (\bibinfo {year}
  {2016})}\BibitemShut {NoStop}%
\bibitem [{\citenamefont {Drescher}\ \emph {et~al.}(2019)\citenamefont
  {Drescher}, \citenamefont {Salmhofer},\ and\ \citenamefont
  {Enss}}]{Drescher2019}%
  \BibitemOpen
  \bibfield  {author} {\bibinfo {author} {\bibfnamefont {M.}~\bibnamefont
  {Drescher}}, \bibinfo {author} {\bibfnamefont {M.}~\bibnamefont
  {Salmhofer}},\ and\ \bibinfo {author} {\bibfnamefont {T.}~\bibnamefont
  {Enss}},\ }\bibfield  {title} {\bibinfo {title} {Real-space dynamics of
  attractive and repulsive polarons in bose-einstein condensates},\ }\href
  {https://doi.org/10.1103/PhysRevA.99.023601} {\bibfield  {journal} {\bibinfo
  {journal} {Phys. Rev. A}\ }\textbf {\bibinfo {volume} {99}},\ \bibinfo
  {pages} {023601} (\bibinfo {year} {2019})}\BibitemShut {NoStop}%
\bibitem [{\citenamefont {Van~Loon}\ \emph {et~al.}(2018)\citenamefont
  {Van~Loon}, \citenamefont {Casteels},\ and\ \citenamefont
  {Tempere}}]{Loon2018}%
  \BibitemOpen
  \bibfield  {author} {\bibinfo {author} {\bibfnamefont {S.}~\bibnamefont
  {Van~Loon}}, \bibinfo {author} {\bibfnamefont {W.}~\bibnamefont {Casteels}},\
  and\ \bibinfo {author} {\bibfnamefont {J.}~\bibnamefont {Tempere}},\
  }\bibfield  {title} {\bibinfo {title} {Ground-state properties of interacting
  bose polarons},\ }\href {https://doi.org/10.1103/PhysRevA.98.063631}
  {\bibfield  {journal} {\bibinfo  {journal} {Phys. Rev. A}\ }\textbf {\bibinfo
  {volume} {98}},\ \bibinfo {pages} {063631} (\bibinfo {year}
  {2018})}\BibitemShut {NoStop}%
\bibitem [{\citenamefont {Rath}\ and\ \citenamefont
  {Schmidt}(2013)}]{Pattrick2013}%
  \BibitemOpen
  \bibfield  {author} {\bibinfo {author} {\bibfnamefont {S.~P.}\ \bibnamefont
  {Rath}}\ and\ \bibinfo {author} {\bibfnamefont {R.}~\bibnamefont {Schmidt}},\
  }\bibfield  {title} {\bibinfo {title} {Field-theoretical study of the bose
  polaron},\ }\href {https://doi.org/10.1103/PhysRevA.88.053632} {\bibfield
  {journal} {\bibinfo  {journal} {Phys. Rev. A}\ }\textbf {\bibinfo {volume}
  {88}},\ \bibinfo {pages} {053632} (\bibinfo {year} {2013})}\BibitemShut
  {NoStop}%
\bibitem [{\citenamefont {Lampo}\ \emph {et~al.}(2018)\citenamefont {Lampo},
  \citenamefont {Charalambous}, \citenamefont {Garc\'{\i}a-March},\ and\
  \citenamefont {Lewenstein}}]{Lampo2018}%
  \BibitemOpen
  \bibfield  {author} {\bibinfo {author} {\bibfnamefont {A.}~\bibnamefont
  {Lampo}}, \bibinfo {author} {\bibfnamefont {C.}~\bibnamefont {Charalambous}},
  \bibinfo {author} {\bibfnamefont {M.~A.}\ \bibnamefont {Garc\'{\i}a-March}},\
  and\ \bibinfo {author} {\bibfnamefont {M.}~\bibnamefont {Lewenstein}},\
  }\bibfield  {title} {\bibinfo {title} {Non-markovian polaron dynamics in a
  trapped bose-einstein condensate},\ }\href
  {https://doi.org/10.1103/PhysRevA.98.063630} {\bibfield  {journal} {\bibinfo
  {journal} {Phys. Rev. A}\ }\textbf {\bibinfo {volume} {98}},\ \bibinfo
  {pages} {063630} (\bibinfo {year} {2018})}\BibitemShut {NoStop}%
\bibitem [{\citenamefont {Khan}\ \emph {et~al.}(2021)\citenamefont {Khan},
  \citenamefont {Ter\ifmmode~\mbox{\c{c}}\else \c{c}\fi{}as}, \citenamefont
  {Mendon\ifmmode~\mbox{\c{c}}\else \c{c}\fi{}a}, \citenamefont {Wehr},
  \citenamefont {Charalambous}, \citenamefont {Lewenstein},\ and\ \citenamefont
  {Garcia-March}}]{Kahn2021}%
  \BibitemOpen
  \bibfield  {author} {\bibinfo {author} {\bibfnamefont {M.~M.}\ \bibnamefont
  {Khan}}, \bibinfo {author} {\bibfnamefont {H.}~\bibnamefont
  {Ter\ifmmode~\mbox{\c{c}}\else \c{c}\fi{}as}}, \bibinfo {author}
  {\bibfnamefont {J.~T.}\ \bibnamefont {Mendon\ifmmode~\mbox{\c{c}}\else
  \c{c}\fi{}a}}, \bibinfo {author} {\bibfnamefont {J.}~\bibnamefont {Wehr}},
  \bibinfo {author} {\bibfnamefont {C.}~\bibnamefont {Charalambous}}, \bibinfo
  {author} {\bibfnamefont {M.}~\bibnamefont {Lewenstein}},\ and\ \bibinfo
  {author} {\bibfnamefont {M.~A.}\ \bibnamefont {Garcia-March}},\ }\bibfield
  {title} {\bibinfo {title} {Quantum dynamics of a bose polaron in a
  $d$-dimensional bose-einstein condensate},\ }\href
  {https://doi.org/10.1103/PhysRevA.103.023303} {\bibfield  {journal} {\bibinfo
   {journal} {Phys. Rev. A}\ }\textbf {\bibinfo {volume} {103}},\ \bibinfo
  {pages} {023303} (\bibinfo {year} {2021})}\BibitemShut {NoStop}%
\bibitem [{\citenamefont {Mistakidis}\ \emph {et~al.}(2019)\citenamefont
  {Mistakidis}, \citenamefont {Katsimiga}, \citenamefont {Koutentakis},
  \citenamefont {Busch},\ and\ \citenamefont {Schmelcher}}]{Mistakidis2019}%
  \BibitemOpen
  \bibfield  {author} {\bibinfo {author} {\bibfnamefont {S.~I.}\ \bibnamefont
  {Mistakidis}}, \bibinfo {author} {\bibfnamefont {G.~C.}\ \bibnamefont
  {Katsimiga}}, \bibinfo {author} {\bibfnamefont {G.~M.}\ \bibnamefont
  {Koutentakis}}, \bibinfo {author} {\bibfnamefont {T.}~\bibnamefont {Busch}},\
  and\ \bibinfo {author} {\bibfnamefont {P.}~\bibnamefont {Schmelcher}},\
  }\bibfield  {title} {\bibinfo {title} {Quench dynamics and orthogonality
  catastrophe of bose polarons},\ }\href
  {https://doi.org/10.1103/PhysRevLett.122.183001} {\bibfield  {journal}
  {\bibinfo  {journal} {Phys. Rev. Lett.}\ }\textbf {\bibinfo {volume} {122}},\
  \bibinfo {pages} {183001} (\bibinfo {year} {2019})}\BibitemShut {NoStop}%
\bibitem [{\citenamefont {Mistakidis}\ \emph {et~al.}(2020)\citenamefont
  {Mistakidis}, \citenamefont {Katsimiga}, \citenamefont {Koutentakis},
  \citenamefont {Busch},\ and\ \citenamefont {Schmelcher}}]{Mistakidis2020}%
  \BibitemOpen
  \bibfield  {author} {\bibinfo {author} {\bibfnamefont {S.~I.}\ \bibnamefont
  {Mistakidis}}, \bibinfo {author} {\bibfnamefont {G.~C.}\ \bibnamefont
  {Katsimiga}}, \bibinfo {author} {\bibfnamefont {G.~M.}\ \bibnamefont
  {Koutentakis}}, \bibinfo {author} {\bibfnamefont {T.}~\bibnamefont {Busch}},\
  and\ \bibinfo {author} {\bibfnamefont {P.}~\bibnamefont {Schmelcher}},\
  }\bibfield  {title} {\bibinfo {title} {Pump-probe spectroscopy of bose
  polarons: Dynamical formation and coherence},\ }\href
  {https://doi.org/10.1103/PhysRevResearch.2.033380} {\bibfield  {journal}
  {\bibinfo  {journal} {Phys. Rev. Research}\ }\textbf {\bibinfo {volume}
  {2}},\ \bibinfo {pages} {033380} (\bibinfo {year} {2020})}\BibitemShut
  {NoStop}%
\bibitem [{\citenamefont {Mistakidis}\ \emph {et~al.}(2021)\citenamefont
  {Mistakidis}, \citenamefont {Koutentakis}, \citenamefont {Grusdt},
  \citenamefont {Sadeghpour},\ and\ \citenamefont
  {Schmelcher}}]{Mistakidis2021}%
  \BibitemOpen
  \bibfield  {author} {\bibinfo {author} {\bibfnamefont {S.~I.}\ \bibnamefont
  {Mistakidis}}, \bibinfo {author} {\bibfnamefont {G.~M.}\ \bibnamefont
  {Koutentakis}}, \bibinfo {author} {\bibfnamefont {F.}~\bibnamefont {Grusdt}},
  \bibinfo {author} {\bibfnamefont {H.~R.}\ \bibnamefont {Sadeghpour}},\ and\
  \bibinfo {author} {\bibfnamefont {P.}~\bibnamefont {Schmelcher}},\ }\bibfield
   {title} {\bibinfo {title} {Radiofrequency spectroscopy of one-dimensional
  trapped bose polarons: crossover from the adiabatic to the diabatic regime},\
  }\href {https://doi.org/10.1088/1367-2630/abe9d5} {\bibfield  {journal}
  {\bibinfo  {journal} {New Journal of Physics}\ }\textbf {\bibinfo {volume}
  {23}},\ \bibinfo {pages} {043051} (\bibinfo {year} {2021})}\BibitemShut
  {NoStop}%
\bibitem [{\citenamefont {J\o{}rgensen}\ \emph {et~al.}(2016)\citenamefont
  {J\o{}rgensen}, \citenamefont {Wacker}, \citenamefont {Skalmstang},
  \citenamefont {Parish}, \citenamefont {Levinsen}, \citenamefont
  {Christensen}, \citenamefont {Bruun},\ and\ \citenamefont
  {Arlt}}]{Jorgensen2016}%
  \BibitemOpen
  \bibfield  {author} {\bibinfo {author} {\bibfnamefont {N.~B.}\ \bibnamefont
  {J\o{}rgensen}}, \bibinfo {author} {\bibfnamefont {L.}~\bibnamefont
  {Wacker}}, \bibinfo {author} {\bibfnamefont {K.~T.}\ \bibnamefont
  {Skalmstang}}, \bibinfo {author} {\bibfnamefont {M.~M.}\ \bibnamefont
  {Parish}}, \bibinfo {author} {\bibfnamefont {J.}~\bibnamefont {Levinsen}},
  \bibinfo {author} {\bibfnamefont {R.~S.}\ \bibnamefont {Christensen}},
  \bibinfo {author} {\bibfnamefont {G.~M.}\ \bibnamefont {Bruun}},\ and\
  \bibinfo {author} {\bibfnamefont {J.~J.}\ \bibnamefont {Arlt}},\ }\bibfield
  {title} {\bibinfo {title} {Observation of attractive and repulsive polarons
  in a bose-einstein condensate},\ }\href
  {https://doi.org/10.1103/PhysRevLett.117.055302} {\bibfield  {journal}
  {\bibinfo  {journal} {Phys. Rev. Lett.}\ }\textbf {\bibinfo {volume} {117}},\
  \bibinfo {pages} {055302} (\bibinfo {year} {2016})}\BibitemShut {NoStop}%
\bibitem [{\citenamefont {Hu}\ \emph {et~al.}(2016)\citenamefont {Hu},
  \citenamefont {Van~de Graaff}, \citenamefont {Kedar}, \citenamefont {Corson},
  \citenamefont {Cornell},\ and\ \citenamefont {Jin}}]{Hu2016}%
  \BibitemOpen
  \bibfield  {author} {\bibinfo {author} {\bibfnamefont {M.-G.}\ \bibnamefont
  {Hu}}, \bibinfo {author} {\bibfnamefont {M.~J.}\ \bibnamefont {Van~de
  Graaff}}, \bibinfo {author} {\bibfnamefont {D.}~\bibnamefont {Kedar}},
  \bibinfo {author} {\bibfnamefont {J.~P.}\ \bibnamefont {Corson}}, \bibinfo
  {author} {\bibfnamefont {E.~A.}\ \bibnamefont {Cornell}},\ and\ \bibinfo
  {author} {\bibfnamefont {D.~S.}\ \bibnamefont {Jin}},\ }\bibfield  {title}
  {\bibinfo {title} {Bose polarons in the strongly interacting regime},\ }\href
  {https://doi.org/10.1103/PhysRevLett.117.055301} {\bibfield  {journal}
  {\bibinfo  {journal} {Phys. Rev. Lett.}\ }\textbf {\bibinfo {volume} {117}},\
  \bibinfo {pages} {055301} (\bibinfo {year} {2016})}\BibitemShut {NoStop}%
\bibitem [{\citenamefont {Levinsen}\ \emph {et~al.}(2017)\citenamefont
  {Levinsen}, \citenamefont {Parish}, \citenamefont {Christensen},
  \citenamefont {Arlt},\ and\ \citenamefont {Bruun}}]{Levinsen2017}%
  \BibitemOpen
  \bibfield  {author} {\bibinfo {author} {\bibfnamefont {J.}~\bibnamefont
  {Levinsen}}, \bibinfo {author} {\bibfnamefont {M.~M.}\ \bibnamefont
  {Parish}}, \bibinfo {author} {\bibfnamefont {R.~S.}\ \bibnamefont
  {Christensen}}, \bibinfo {author} {\bibfnamefont {J.~J.}\ \bibnamefont
  {Arlt}},\ and\ \bibinfo {author} {\bibfnamefont {G.~M.}\ \bibnamefont
  {Bruun}},\ }\bibfield  {title} {\bibinfo {title} {Finite-temperature behavior
  of the bose polaron},\ }\href {https://doi.org/10.1103/PhysRevA.96.063622}
  {\bibfield  {journal} {\bibinfo  {journal} {Phys. Rev. A}\ }\textbf {\bibinfo
  {volume} {96}},\ \bibinfo {pages} {063622} (\bibinfo {year}
  {2017})}\BibitemShut {NoStop}%
\bibitem [{\citenamefont {Boudjem\^aa}(2014)}]{Boudjemaa2014_I}%
  \BibitemOpen
  \bibfield  {author} {\bibinfo {author} {\bibfnamefont {A.}~\bibnamefont
  {Boudjem\^aa}},\ }\bibfield  {title} {\bibinfo {title} {Self-localized state
  and solitons in a bose-einstein-condensate--impurity mixture at finite
  temperature},\ }\href {https://doi.org/10.1103/PhysRevA.90.013628} {\bibfield
   {journal} {\bibinfo  {journal} {Phys. Rev. A}\ }\textbf {\bibinfo {volume}
  {90}},\ \bibinfo {pages} {013628} (\bibinfo {year} {2014})}\BibitemShut
  {NoStop}%
\bibitem [{\citenamefont {Boudjem{\^{a}}a}(2014)}]{Boudjemaa2014_II}%
  \BibitemOpen
  \bibfield  {author} {\bibinfo {author} {\bibfnamefont {A.}~\bibnamefont
  {Boudjem{\^{a}}a}},\ }\bibfield  {title} {\bibinfo {title} {Self-consistent
  theory of a bose{\textendash}einstein condensate with impurity at finite
  temperature},\ }\href {https://doi.org/10.1088/1751-8113/48/4/045002}
  {\bibfield  {journal} {\bibinfo  {journal} {Journal of Physics A:
  Mathematical and Theoretical}\ }\textbf {\bibinfo {volume} {48}},\ \bibinfo
  {pages} {045002} (\bibinfo {year} {2014})}\BibitemShut {NoStop}%
\bibitem [{\citenamefont {Pastukhov}(2018)}]{Pastukhov2018}%
  \BibitemOpen
  \bibfield  {author} {\bibinfo {author} {\bibfnamefont {V.}~\bibnamefont
  {Pastukhov}},\ }\bibfield  {title} {\bibinfo {title} {Polaron in the dilute
  critical bose condensate},\ }\href {https://doi.org/10.1088/1751-8121/aab9c1}
  {\bibfield  {journal} {\bibinfo  {journal} {Journal of Physics A:
  Mathematical and Theoretical}\ }\textbf {\bibinfo {volume} {51}},\ \bibinfo
  {pages} {195003} (\bibinfo {year} {2018})}\BibitemShut {NoStop}%
\bibitem [{\citenamefont {Guenther}\ \emph {et~al.}(2018)\citenamefont
  {Guenther}, \citenamefont {Massignan}, \citenamefont {Lewenstein},\ and\
  \citenamefont {Bruun}}]{Guenther2018}%
  \BibitemOpen
  \bibfield  {author} {\bibinfo {author} {\bibfnamefont {N.-E.}\ \bibnamefont
  {Guenther}}, \bibinfo {author} {\bibfnamefont {P.}~\bibnamefont {Massignan}},
  \bibinfo {author} {\bibfnamefont {M.}~\bibnamefont {Lewenstein}},\ and\
  \bibinfo {author} {\bibfnamefont {G.~M.}\ \bibnamefont {Bruun}},\ }\bibfield
  {title} {\bibinfo {title} {Bose polarons at finite temperature and strong
  coupling},\ }\href {https://doi.org/10.1103/PhysRevLett.120.050405}
  {\bibfield  {journal} {\bibinfo  {journal} {Phys. Rev. Lett.}\ }\textbf
  {\bibinfo {volume} {120}},\ \bibinfo {pages} {050405} (\bibinfo {year}
  {2018})}\BibitemShut {NoStop}%
\bibitem [{\citenamefont {Field}\ \emph {et~al.}(2020)\citenamefont {Field},
  \citenamefont {Levinsen},\ and\ \citenamefont {Parish}}]{Field2020}%
  \BibitemOpen
  \bibfield  {author} {\bibinfo {author} {\bibfnamefont {B.}~\bibnamefont
  {Field}}, \bibinfo {author} {\bibfnamefont {J.}~\bibnamefont {Levinsen}},\
  and\ \bibinfo {author} {\bibfnamefont {M.~M.}\ \bibnamefont {Parish}},\
  }\bibfield  {title} {\bibinfo {title} {Fate of the bose polaron at finite
  temperature},\ }\href {https://doi.org/10.1103/PhysRevA.101.013623}
  {\bibfield  {journal} {\bibinfo  {journal} {Phys. Rev. A}\ }\textbf {\bibinfo
  {volume} {101}},\ \bibinfo {pages} {013623} (\bibinfo {year}
  {2020})}\BibitemShut {NoStop}%
\bibitem [{\citenamefont {Dzsotjan}\ \emph {et~al.}(2020)\citenamefont
  {Dzsotjan}, \citenamefont {Schmidt},\ and\ \citenamefont
  {Fleischhauer}}]{Dzsotjan2020}%
  \BibitemOpen
  \bibfield  {author} {\bibinfo {author} {\bibfnamefont {D.}~\bibnamefont
  {Dzsotjan}}, \bibinfo {author} {\bibfnamefont {R.}~\bibnamefont {Schmidt}},\
  and\ \bibinfo {author} {\bibfnamefont {M.}~\bibnamefont {Fleischhauer}},\
  }\bibfield  {title} {\bibinfo {title} {Dynamical variational approach to bose
  polarons at finite temperatures},\ }\href
  {https://doi.org/10.1103/PhysRevLett.124.223401} {\bibfield  {journal}
  {\bibinfo  {journal} {Phys. Rev. Lett.}\ }\textbf {\bibinfo {volume} {124}},\
  \bibinfo {pages} {223401} (\bibinfo {year} {2020})}\BibitemShut {NoStop}%
\bibitem [{\citenamefont {Yan}\ \emph {et~al.}(2020)\citenamefont {Yan},
  \citenamefont {Ni}, \citenamefont {Robens},\ and\ \citenamefont
  {Zwierlein}}]{Yan2020}%
  \BibitemOpen
  \bibfield  {author} {\bibinfo {author} {\bibfnamefont {Z.~Z.}\ \bibnamefont
  {Yan}}, \bibinfo {author} {\bibfnamefont {Y.}~\bibnamefont {Ni}}, \bibinfo
  {author} {\bibfnamefont {C.}~\bibnamefont {Robens}},\ and\ \bibinfo {author}
  {\bibfnamefont {M.~W.}\ \bibnamefont {Zwierlein}},\ }\bibfield  {title}
  {\bibinfo {title} {Bose polarons near quantum criticality},\ }\href
  {https://doi.org/10.1126/science.aax5850} {\bibfield  {journal} {\bibinfo
  {journal} {Science}\ }\textbf {\bibinfo {volume} {368}},\ \bibinfo {pages}
  {190} (\bibinfo {year} {2020})}\BibitemShut {NoStop}%
\bibitem [{\citenamefont {Sachdev}(2011)}]{Sachdev2011}%
  \BibitemOpen
  \bibfield  {author} {\bibinfo {author} {\bibfnamefont {S.}~\bibnamefont
  {Sachdev}},\ }\href {https://doi.org/10.1017/CBO9780511973765} {\emph
  {\bibinfo {title} {Quantum Phase Transitions}}},\ \bibinfo {edition} {2nd}\
  ed.\ (\bibinfo  {publisher} {Cambridge University Press},\ \bibinfo {year}
  {2011})\BibitemShut {NoStop}%
\bibitem [{\citenamefont {Lee}\ \emph {et~al.}(2006)\citenamefont {Lee},
  \citenamefont {Nagaosa},\ and\ \citenamefont {Wen}}]{Lee2006}%
  \BibitemOpen
  \bibfield  {author} {\bibinfo {author} {\bibfnamefont {P.~A.}\ \bibnamefont
  {Lee}}, \bibinfo {author} {\bibfnamefont {N.}~\bibnamefont {Nagaosa}},\ and\
  \bibinfo {author} {\bibfnamefont {X.-G.}\ \bibnamefont {Wen}},\ }\bibfield
  {title} {\bibinfo {title} {Doping a mott insulator: Physics of
  high-temperature superconductivity},\ }\href
  {https://doi.org/10.1103/RevModPhys.78.17} {\bibfield  {journal} {\bibinfo
  {journal} {Rev. Mod. Phys.}\ }\textbf {\bibinfo {volume} {78}},\ \bibinfo
  {pages} {17} (\bibinfo {year} {2006})}\BibitemShut {NoStop}%
\bibitem [{\citenamefont {Hartnoll}(2015)}]{Hartnoll2015}%
  \BibitemOpen
  \bibfield  {author} {\bibinfo {author} {\bibfnamefont {S.~A.}\ \bibnamefont
  {Hartnoll}},\ }\bibfield  {title} {\bibinfo {title} {Theory of universal
  incoherent metallic transport},\ }\href {https://doi.org/10.1038/nphys3174}
  {\bibfield  {journal} {\bibinfo  {journal} {Nature Physics}\ }\textbf
  {\bibinfo {volume} {11}},\ \bibinfo {pages} {54} (\bibinfo {year}
  {2015})}\BibitemShut {NoStop}%
\bibitem [{Sup()}]{Supplement}%
  \BibitemOpen
  \href@noop {} {\bibinfo {title} {See {S}upplemental material which includes
  {R}efs.
  \cite{Chin2002,Thesis_Rota,Boninsegni2006,Landau1977,Pilati2006,Flugge1994,Sarsa2000,Galli2003,Cuervo2006,Rota2010}
  for more details on {PIMC} and {PIGS} methods and for a discussion on the
  interacting potentials used in this work.}}\BibitemShut {Stop}%
\bibitem [{\citenamefont {Giorgini}\ \emph {et~al.}(1999)\citenamefont
  {Giorgini}, \citenamefont {Boronat},\ and\ \citenamefont
  {Casulleras}}]{Giorgini1999}%
  \BibitemOpen
  \bibfield  {author} {\bibinfo {author} {\bibfnamefont {S.}~\bibnamefont
  {Giorgini}}, \bibinfo {author} {\bibfnamefont {J.}~\bibnamefont {Boronat}},\
  and\ \bibinfo {author} {\bibfnamefont {J.}~\bibnamefont {Casulleras}},\
  }\bibfield  {title} {\bibinfo {title} {Ground state of a homogeneous bose
  gas: A diffusion monte carlo calculation},\ }\href
  {https://doi.org/10.1103/PhysRevA.60.5129} {\bibfield  {journal} {\bibinfo
  {journal} {Phys. Rev. A}\ }\textbf {\bibinfo {volume} {60}},\ \bibinfo
  {pages} {5129} (\bibinfo {year} {1999})}\BibitemShut {NoStop}%
\bibitem [{\citenamefont {Sakkos}\ \emph {et~al.}(2009)\citenamefont {Sakkos},
  \citenamefont {Casulleras},\ and\ \citenamefont {Boronat}}]{Sakkos2009}%
  \BibitemOpen
  \bibfield  {author} {\bibinfo {author} {\bibfnamefont {K.}~\bibnamefont
  {Sakkos}}, \bibinfo {author} {\bibfnamefont {J.}~\bibnamefont {Casulleras}},\
  and\ \bibinfo {author} {\bibfnamefont {J.}~\bibnamefont {Boronat}},\
  }\bibfield  {title} {\bibinfo {title} {High order chin actions in path
  integral monte carlo},\ }\href {https://doi.org/10.1063/1.3143522} {\bibfield
   {journal} {\bibinfo  {journal} {The Journal of Chemical Physics}\ }\textbf
  {\bibinfo {volume} {130}},\ \bibinfo {pages} {204109} (\bibinfo {year}
  {2009})}\BibitemShut {NoStop}%
\bibitem [{\citenamefont {Boninsegni}\ and\ \citenamefont
  {Ceperley}(1995)}]{Boninsegni1995}%
  \BibitemOpen
  \bibfield  {author} {\bibinfo {author} {\bibfnamefont {M.}~\bibnamefont
  {Boninsegni}}\ and\ \bibinfo {author} {\bibfnamefont {D.~M.}\ \bibnamefont
  {Ceperley}},\ }\bibfield  {title} {\bibinfo {title} {Path integral monte
  carlo simulation of isotopic liquid helium mixtures},\ }\href
  {https://doi.org/10.1103/PhysRevLett.74.2288} {\bibfield  {journal} {\bibinfo
   {journal} {Phys. Rev. Lett.}\ }\textbf {\bibinfo {volume} {74}},\ \bibinfo
  {pages} {2288} (\bibinfo {year} {1995})}\BibitemShut {NoStop}%
\bibitem [{\citenamefont {Boronat}\ and\ \citenamefont
  {Casulleras}(1999)}]{Boronat1999}%
  \BibitemOpen
  \bibfield  {author} {\bibinfo {author} {\bibfnamefont {J.}~\bibnamefont
  {Boronat}}\ and\ \bibinfo {author} {\bibfnamefont {J.}~\bibnamefont
  {Casulleras}},\ }\bibfield  {title} {\bibinfo {title} {Quantum monte carlo
  study of static properties of one ${}^{3}\mathrm{He}$ atom in superfluid
  ${}^{4}\mathrm{He}$},\ }\href {https://doi.org/10.1103/PhysRevB.59.8844}
  {\bibfield  {journal} {\bibinfo  {journal} {Phys. Rev. B}\ }\textbf {\bibinfo
  {volume} {59}},\ \bibinfo {pages} {8844} (\bibinfo {year}
  {1999})}\BibitemShut {NoStop}%
\bibitem [{\citenamefont {Ceperley}(1995)}]{Ceperley1995}%
  \BibitemOpen
  \bibfield  {author} {\bibinfo {author} {\bibfnamefont {D.~M.}\ \bibnamefont
  {Ceperley}},\ }\bibfield  {title} {\bibinfo {title} {Path integrals in the
  theory of condensed helium},\ }\href
  {https://doi.org/10.1103/RevModPhys.67.279} {\bibfield  {journal} {\bibinfo
  {journal} {Rev. Mod. Phys.}\ }\textbf {\bibinfo {volume} {67}},\ \bibinfo
  {pages} {279} (\bibinfo {year} {1995})}\BibitemShut {NoStop}%
\bibitem [{\citenamefont {Ferr\'e}\ and\ \citenamefont
  {Boronat}(2016)}]{Ferre2016}%
  \BibitemOpen
  \bibfield  {author} {\bibinfo {author} {\bibfnamefont {G.}~\bibnamefont
  {Ferr\'e}}\ and\ \bibinfo {author} {\bibfnamefont {J.}~\bibnamefont
  {Boronat}},\ }\bibfield  {title} {\bibinfo {title} {Dynamic structure factor
  of liquid $^{4}\mathrm{He}$ across the normal-superfluid transition},\ }\href
  {https://doi.org/10.1103/PhysRevB.93.104510} {\bibfield  {journal} {\bibinfo
  {journal} {Phys. Rev. B}\ }\textbf {\bibinfo {volume} {93}},\ \bibinfo
  {pages} {104510} (\bibinfo {year} {2016})}\BibitemShut {NoStop}%
\bibitem [{\citenamefont {MacKeow}(1997)}]{MacKeow1997}%
  \BibitemOpen
  \bibfield  {author} {\bibinfo {author} {\bibfnamefont {P.~K.}\ \bibnamefont
  {MacKeow}},\ }\href@noop {} {\emph {\bibinfo {title} {Stochastic Simulation
  in Physics}}}\ (\bibinfo  {publisher} {Springer},\ \bibinfo {year}
  {1997})\BibitemShut {NoStop}%
\bibitem [{\citenamefont {Fabrocini}\ \emph {et~al.}(1986)\citenamefont
  {Fabrocini}, \citenamefont {Fantoni}, \citenamefont {Rosati},\ and\
  \citenamefont {Polls}}]{Fabrocini1986}%
  \BibitemOpen
  \bibfield  {author} {\bibinfo {author} {\bibfnamefont {A.}~\bibnamefont
  {Fabrocini}}, \bibinfo {author} {\bibfnamefont {S.}~\bibnamefont {Fantoni}},
  \bibinfo {author} {\bibfnamefont {S.}~\bibnamefont {Rosati}},\ and\ \bibinfo
  {author} {\bibfnamefont {A.}~\bibnamefont {Polls}},\ }\bibfield  {title}
  {\bibinfo {title} {Microscopic calculations of the excitation spectrum of one
  $^{3}\mathrm{He}$ impurity in liquid $^{4}\mathrm{He}$},\ }\href
  {https://doi.org/10.1103/PhysRevB.33.6057} {\bibfield  {journal} {\bibinfo
  {journal} {Phys. Rev. B}\ }\textbf {\bibinfo {volume} {33}},\ \bibinfo
  {pages} {6057} (\bibinfo {year} {1986})}\BibitemShut {NoStop}%
\bibitem [{\citenamefont {Fabrocini}\ and\ \citenamefont
  {Polls}(1998)}]{Fabrocini1998}%
  \BibitemOpen
  \bibfield  {author} {\bibinfo {author} {\bibfnamefont {A.}~\bibnamefont
  {Fabrocini}}\ and\ \bibinfo {author} {\bibfnamefont {A.}~\bibnamefont
  {Polls}},\ }\bibfield  {title} {\bibinfo {title} {${}^{3}\mathrm{He}$
  impurity excitation spectrum in liquid ${}^{4}\mathrm{He}$},\ }\href
  {https://doi.org/10.1103/PhysRevB.58.5209} {\bibfield  {journal} {\bibinfo
  {journal} {Phys. Rev. B}\ }\textbf {\bibinfo {volume} {58}},\ \bibinfo
  {pages} {5209} (\bibinfo {year} {1998})}\BibitemShut {NoStop}%
\bibitem [{\citenamefont {Chin}\ and\ \citenamefont {Chen}(2002)}]{Chin2002}%
  \BibitemOpen
  \bibfield  {author} {\bibinfo {author} {\bibfnamefont {S.~A.}\ \bibnamefont
  {Chin}}\ and\ \bibinfo {author} {\bibfnamefont {C.~R.}\ \bibnamefont
  {Chen}},\ }\bibfield  {title} {\bibinfo {title} {Gradient symplectic
  algorithms for solving the schrödinger equation with time-dependent
  potentials},\ }\href {https://doi.org/10.1063/1.1485725} {\bibfield
  {journal} {\bibinfo  {journal} {The Journal of Chemical Physics}\ }\textbf
  {\bibinfo {volume} {117}},\ \bibinfo {pages} {1409} (\bibinfo {year}
  {2002})}\BibitemShut {NoStop}%
\bibitem [{\citenamefont {Rota}(2011)}]{Thesis_Rota}%
  \BibitemOpen
  \bibfield  {author} {\bibinfo {author} {\bibfnamefont {R.}~\bibnamefont
  {Rota}},\ }\emph {\bibinfo {title} {Path Integral Monte Carlo and
  Bose-Einstein condensation in quantum fluids and solids}},\ \href@noop {}
  {Ph.D. thesis},\ \bibinfo  {school} {Universitat Politècnica de Catalunya}
  (\bibinfo {year} {2011})\BibitemShut {NoStop}%
\bibitem [{\citenamefont {Boninsegni}\ \emph {et~al.}(2006)\citenamefont
  {Boninsegni}, \citenamefont {Prokof'ev},\ and\ \citenamefont
  {Svistunov}}]{Boninsegni2006}%
  \BibitemOpen
  \bibfield  {author} {\bibinfo {author} {\bibfnamefont {M.}~\bibnamefont
  {Boninsegni}}, \bibinfo {author} {\bibfnamefont {N.~V.}\ \bibnamefont
  {Prokof'ev}},\ and\ \bibinfo {author} {\bibfnamefont {B.~V.}\ \bibnamefont
  {Svistunov}},\ }\bibfield  {title} {\bibinfo {title} {Worm algorithm and
  diagrammatic {M}onte {C}arlo: A new approach to continuous-space path
  integral {M}onte {C}arlo simulations},\ }\href
  {https://doi.org/10.1103/PhysRevE.74.036701} {\bibfield  {journal} {\bibinfo
  {journal} {Phys. Rev. E}\ }\textbf {\bibinfo {volume} {74}},\ \bibinfo
  {pages} {036701} (\bibinfo {year} {2006})}\BibitemShut {NoStop}%
\bibitem [{\citenamefont {Landau}\ and\ \citenamefont
  {Lifshitz}(1977)}]{Landau1977}%
  \BibitemOpen
  \bibfield  {author} {\bibinfo {author} {\bibfnamefont {L.~D.}\ \bibnamefont
  {Landau}}\ and\ \bibinfo {author} {\bibfnamefont {E.~M.}\ \bibnamefont
  {Lifshitz}},\ }\href@noop {} {\emph {\bibinfo {title} {Quantum Mechanics
  (Non-relativistic Theory)}}}\ (\bibinfo  {publisher} {Pergamon Press},\
  \bibinfo {address} {Oxford},\ \bibinfo {year} {1977})\ p.\ \bibinfo {pages}
  {550}\BibitemShut {NoStop}%
\bibitem [{\citenamefont {Pilati}\ \emph {et~al.}(2006)\citenamefont {Pilati},
  \citenamefont {Sakkos}, \citenamefont {Boronat}, \citenamefont {Casulleras},\
  and\ \citenamefont {Giorgini}}]{Pilati2006}%
  \BibitemOpen
  \bibfield  {author} {\bibinfo {author} {\bibfnamefont {S.}~\bibnamefont
  {Pilati}}, \bibinfo {author} {\bibfnamefont {K.}~\bibnamefont {Sakkos}},
  \bibinfo {author} {\bibfnamefont {J.}~\bibnamefont {Boronat}}, \bibinfo
  {author} {\bibfnamefont {J.}~\bibnamefont {Casulleras}},\ and\ \bibinfo
  {author} {\bibfnamefont {S.}~\bibnamefont {Giorgini}},\ }\bibfield  {title}
  {\bibinfo {title} {Equation of state of an interacting bose gas at finite
  temperature: A path-integral monte carlo study},\ }\href
  {https://doi.org/10.1103/PhysRevA.74.043621} {\bibfield  {journal} {\bibinfo
  {journal} {Phys. Rev. A}\ }\textbf {\bibinfo {volume} {74}},\ \bibinfo
  {pages} {043621} (\bibinfo {year} {2006})}\BibitemShut {NoStop}%
\bibitem [{\citenamefont {Fl{\"u}gge}(1994)}]{Flugge1994}%
  \BibitemOpen
  \bibfield  {author} {\bibinfo {author} {\bibfnamefont {S.}~\bibnamefont
  {Fl{\"u}gge}},\ }\href@noop {} {\emph {\bibinfo {title} {Practical Quantum
  Mechanics}}}\ (\bibinfo  {publisher} {Springer},\ \bibinfo {address}
  {Germany},\ \bibinfo {year} {1994})\BibitemShut {NoStop}%
\bibitem [{\citenamefont {Sarsa}\ \emph {et~al.}(2000)\citenamefont {Sarsa},
  \citenamefont {Schmidt},\ and\ \citenamefont {Magro}}]{Sarsa2000}%
  \BibitemOpen
  \bibfield  {author} {\bibinfo {author} {\bibfnamefont {A.}~\bibnamefont
  {Sarsa}}, \bibinfo {author} {\bibfnamefont {K.~E.}\ \bibnamefont {Schmidt}},\
  and\ \bibinfo {author} {\bibfnamefont {W.~R.}\ \bibnamefont {Magro}},\
  }\bibfield  {title} {\bibinfo {title} {A path integral ground state method},\
  }\href {https://doi.org/10.1063/1.481926} {\bibfield  {journal} {\bibinfo
  {journal} {The Journal of Chemical Physics}\ }\textbf {\bibinfo {volume}
  {113}},\ \bibinfo {pages} {1366} (\bibinfo {year} {2000})}\BibitemShut
  {NoStop}%
\bibitem [{\citenamefont {Galli}\ and\ \citenamefont
  {Reatto}(2003)}]{Galli2003}%
  \BibitemOpen
  \bibfield  {author} {\bibinfo {author} {\bibfnamefont {D.~E.}\ \bibnamefont
  {Galli}}\ and\ \bibinfo {author} {\bibfnamefont {L.}~\bibnamefont {Reatto}},\
  }\bibfield  {title} {\bibinfo {title} {Recent progress in simulation of the
  ground state of many boson systems},\ }\href
  {https://doi.org/10.1080/0026897031000074562} {\bibfield  {journal} {\bibinfo
   {journal} {Molecular Physics}\ }\textbf {\bibinfo {volume} {101}},\ \bibinfo
  {pages} {1697} (\bibinfo {year} {2003})}\BibitemShut {NoStop}%
\bibitem [{\citenamefont {Cuervo}\ and\ \citenamefont
  {Roy}(2006)}]{Cuervo2006}%
  \BibitemOpen
  \bibfield  {author} {\bibinfo {author} {\bibfnamefont {J.~E.}\ \bibnamefont
  {Cuervo}}\ and\ \bibinfo {author} {\bibfnamefont {P.-N.}\ \bibnamefont
  {Roy}},\ }\bibfield  {title} {\bibinfo {title} {Path integral ground state
  study of finite-size systems: Application to small $(parahydrogen)_{N}
  ({N}=2–20)$ clusters},\ }\href {https://doi.org/10.1063/1.2352735}
  {\bibfield  {journal} {\bibinfo  {journal} {The Journal of Chemical Physics}\
  }\textbf {\bibinfo {volume} {125}},\ \bibinfo {pages} {124314} (\bibinfo
  {year} {2006})}\BibitemShut {NoStop}%
\bibitem [{\citenamefont {Rota}\ \emph {et~al.}(2010)\citenamefont {Rota},
  \citenamefont {Casulleras}, \citenamefont {Mazzanti},\ and\ \citenamefont
  {Boronat}}]{Rota2010}%
  \BibitemOpen
  \bibfield  {author} {\bibinfo {author} {\bibfnamefont {R.}~\bibnamefont
  {Rota}}, \bibinfo {author} {\bibfnamefont {J.}~\bibnamefont {Casulleras}},
  \bibinfo {author} {\bibfnamefont {F.}~\bibnamefont {Mazzanti}},\ and\
  \bibinfo {author} {\bibfnamefont {J.}~\bibnamefont {Boronat}},\ }\bibfield
  {title} {\bibinfo {title} {High-order time expansion path integral ground
  state},\ }\href {https://doi.org/10.1103/PhysRevE.81.016707} {\bibfield
  {journal} {\bibinfo  {journal} {Phys. Rev. E}\ }\textbf {\bibinfo {volume}
  {81}},\ \bibinfo {pages} {016707} (\bibinfo {year} {2010})}\BibitemShut
  {NoStop}%
\end{thebibliography}%

\end{document}